\documentclass[aps,showpacs,prd,onecolumn]{revtex4-2}
\usepackage{eurosym}
\usepackage{amsfonts}
\usepackage{amssymb}
\usepackage{amsmath}
\usepackage{dsfont}
\usepackage{color}
\usepackage[usenames,dvipsnames]{xcolor}
\usepackage{float}
\usepackage{accents}
\usepackage{graphicx}
\usepackage{epstopdf}
\usepackage[colorlinks=true,pdfstartview=FitV,linkcolor=blue,citecolor=blue,urlcolor=blue,breaklinks=true]{hyperref}
\usepackage{ulem}

\begin{document}

\title{Self-dual compact gauged baby skyrmions in a continuous medium}
\author{C. A. I. Flori\'an}
\email{cesar.aif@discente.ufma.br}
\author{Rodolfo Casana}
\email{rodolfo.casana@ufma.br}
\email{rodolfo.casana@gmail.com}
\author{Andr\'e C. Santos}
\email{andre.cs@discente.ufma.br}
\email{andre$\_$cavs@hotmail.com}
\affiliation{Departamento de F\'{\i}sica, Universidade Federal do Maranh\~{a}o,
65080-805, S\~{a}o Lu\'{\i}s, Maranh\~{a}o, Brazil.}

\begin{abstract}
We investigate the existence of self-dual configurations in the restricted
gauged baby Skyrme model enlarged with a $\mathds{Z}_2$--symmetry, which
introduces a real scalar field. For such a purpose, we implement the
Bogomol'nyi procedure that provides a lower bound for the energy and the
respective self-dual equations whose solutions saturate such a bound. Aiming
to solve the self-dual equations, we specifically focused on a class of
topological structures called compacton. We obtain the corresponding
numerical solutions within two distinct scenarios, each defined by a scalar
field, allowing us to describe different magnetic media. Finally, we analyze
how the compacton profiles change when immersed in each medium.
\end{abstract}

\maketitle

\section{Introduction}

The Skyrme model \cite{Skyrme} is a (3+1)-dimensional nonlinear field theory
conceived initially to study some nonperturbative aspects of quantum
Chromodynamics (QCD); for example, it provides physical properties of
hadrons and nuclei \cite{Adkins_2} compatible with the low-energy regimen of
QCD. The hadrons described by the Skyrme model emerge as topological
solitons, so-called skyrmions. The (2+1)-dimensional or planar version \cite%
{Piette}, known as the baby Skyrme model, is a laboratory to study many
aspects of the original Skyrme model. Like the original version, the baby
Skyrme model consists of a $O(3)$ nonlinear sigma term, the Skyrme one, and
a nonlinear potential, $V(\hat{n}\cdot \vec{\phi})$, stabilizing the
solitons \cite{Hobart_10}. The Skyrme field $\vec{\phi}$ is a unit
three-component vector of scalar fields $\vec{\phi} =\left(\phi_{1},
\phi_{2}, \phi _{3} \right)$ obeying $\vec{\phi} \cdot\vec{\phi}=1$, so
describing the unit sphere $\mathbb{S}^{2}$. The unitary vector $\hat{n}$
gives a preferential direction in the internal space $\mathbb{S}^{2}$ \cite%
{fnote}. The Skyrme model is invariant under the $SO(3)$ symmetry, while the
potential partially breaks this symmetry, keeping the $SO(2)$ symmetry
unchanged (which is isomorphic to the $U(1)$ one). Further, the potential
must satisfy the condition $V\left( \phi _{n}\right) \rightarrow 0$ when $%
\phi_{n}\rightarrow 1$, known as the single vacuum configuration. Although
the baby Skyrme model describes stable solitons, it does not support a
Bogomol'nyi-Prasad-Sommerfield (BPS) structure \cite{Bogomol}, i.e., an
energy lower-bound (the Bogomol'nyi bound) and a set of self-dual equations
providing solitons that saturate this energy bound. On the other hand, in
the absence of the sigma term arises the so-called restricted baby Skyrme
model \cite{Gisiger}, which supports BPS or self-dual configurations \cite%
{Adam085007}.

A natural physical extension for the Skyrme model is to couple it with a $%
U(1)$ gauge field \cite{EWitten}, which allows us to investigate the
electric and magnetic properties. In this context, Ref. \cite{Adam_12}
provides the first study showing that the restricted gauge baby Skyrme model
supports a BPS structure. Afterward, BPS or self-dual solutions carrying
both magnetic flux and electric charge were obtained \cite{AdamJhep,
Casana_16, Casana_18}. Such studies in the restricted gauge baby Skyrme
model have shown different BPS soliton profiles, i.e., it supports compact
and noncompact topological structures. Concerning the compact skyrmions,
Ref. \cite{Gisiger} provides the first investigation checking their
existence into the restricted baby Skyrme model (nongauged version).
Further, studies considering the sigma term as $\epsilon \partial_\mu \vec{%
\phi} \cdot \partial^\mu \vec{\phi}$ \cite{r5,r6} show that such a model in
the limit $\epsilon \rightarrow 0$ reproduces exactly the BPS structure of
the restricted baby Skyrme model obtained in the Ref. \cite{Adam085007}.
Posteriorly, studies of the gauged version revealed that it also engenders
compactons \cite{Adam_105013} endowed with magnetic flux, but electrically
neutral \cite{Adam_12}. Moreover, recently, compactons carrying both the
magnetic flux and electric charge were also found \cite{Casana_16}.
Regarding the restricted gauged baby Skyrme model, its BPS structure is
related to the $N=2$ SUSY gauged Skyrme model in $(2+1)$--dimensions, whose
bosonic Lagrangian requires the disappearance of the term $D_\mu \vec{\phi}%
\cdot D^\mu \vec{\phi}$, as established in Refs. \cite{Queiruga0}.

It is well-known that gauge models can be enlarged or extended by including
a new symmetry, for example, in the Maxwell-Higgs model \cite{new_ref}.
Within that context, new topological defects by adding a $\mathds{Z}_{2}$%
--symmetry recently have been founded in the Maxwell-Higgs models \cite%
{Bazeia_780, Bazeia_17}, magnetic monopoles \cite{Bazeia_105024}, gauged $%
CP(2)$ model \cite{Andrade_056014}, and gauged $O(3)$ sigma model \cite%
{Casana_056014}. Remarkably, this additional symmetry can engender a
magnetic or dielectric medium that could profoundly affect the solitons'
physical properties.

On the other hand, the so-called magnetic skyrmions have attracted the
community's attention because they come up in the description of various
physical systems as, e.g., in condensed matter \cite{Volovik_3}, in the
quantum Hall effect \cite{Sondhi_4}, superconductors \cite{Zuyzin_7}, and
magnetic materials \cite{Binz_6}, including recent investigations with the
Dzyaloshinskii-Moriya (DM) interaction \cite{Dohi,BJSchroers}. Furthermore,
in the condensed matter context, effects induced on the topological
structures due to geometrical constraints in the magnetic materials at the
nanometric scale are reported, i.e., the geometry's influence on the
formation of magnetic skyrmions \cite{Zhou, Jiang, Fert, Schaffer}, there
are also among them those with compact support that are very sensitive to
geometrical constraints \cite{Motohiko, Chen, Mantel, Khoo}.

Taking the ideas discussed in the previously, we seek new topological
configurations in limited planar regions simulating geometrical
restrictions. For this purpose, we introduce an extended version of the
gauged BPS baby Skyrme model to study solitonic solutions, focusing on
compactons immersed in a magnetic medium \cite{fnote1}. Thus, the original
symmetry $SO(3)$ is enlarged to $SO(3)\times \mathds{Z}_{2}$, being the $%
\mathds{Z}_{2}$--symmetry that rules a neutral scalar field coupled to the
Abelian gauge field through a magnetic permeability that is a function of
the scalar field itself. Consequently, in our case, including the $\mathds{Z}%
_{2}$--symmetry allows us to modify the compacton profiles, whereas we keep
them in geometrically constrained regions.

We present the results in the following way: In Sec. \ref{SecII}, we
introduce our model and show the corresponding Euler-Lagrange equations.
Besides, we found the BPS structure of the system and solved the respective
BPS equations by considering rotationally symmetric configurations. In Sec. %
\ref{SecVI}, we study two scenarios describing different magnetic media and
solve the BPS system numerically; then, we highlight the new features
presented by the new compacton profiles. Lastly, we make our final remarks
and conclusions in Sec. \ref{SecV}.

\section{The restricted gauged baby Skyrme model in a magnetic medium}

\label{SecII}

We investigate an extension of the restricted baby Skyrme model immersed in
a magnetic medium driven by a real scalar field defined by the Lagrangian
\begin{equation}
L=E_{0}\int d^{2}\mathbf{x}\mathcal{L}\text{,}  \label{2}
\end{equation}%
where $E_{0}$ is a common factor that sets the energy scale (which hereafter
we will take as $E_{0}=1$) and $\mathcal{L}$ is the Lagrangian density \cite%
{metrica} given by
\begin{eqnarray}
\mathcal{L} &=&-\frac{1}{4g^{2}}\Sigma (\chi )F_{\mu \nu }F^{\mu \nu }-\frac{%
\lambda ^{2}}{4}(D_{\mu }\vec{\phi}\times D_{\nu }\vec{\phi})^{2}+\frac{1}{2}%
\partial _{\mu }\chi \partial ^{\mu }\chi -V\left( \phi _{n},\chi \right)
\text{,}  \label{3}
\end{eqnarray}%
with the non-negative real function $\Sigma (\chi )$ setting the
electromagnetic properties of the medium as a function of the real scalar
field $\chi$ introduced to incorporate the $\mathds{Z}_2$--symmetry. $F_{\mu
\nu }=\partial_{\mu }A_{\nu }-\partial _{\nu }A_{\mu }$ is the
strength-tensor of the Abelian gauge field $A_{\mu }$, $g$ is the
electromagnetic coupling constant, and $\lambda $ the Skyrme coupling
constant. In addition, the Skyrme and gauge fields are coupled minimally
through the covariant derivative
\begin{equation}
D_{\mu }\vec{\phi}=\partial _{\mu }\vec{\phi}+A_{\mu }\hat{n}\times \vec{\phi%
}.  \label{4}
\end{equation}%
The potential ${V}\left(\phi _{n}, \chi \right) $ stands for some
appropriated interaction in terms of both the Skyrme and the scalar fields.
Furthermore, the scalar fields $\vec{\phi}$ and $\chi $ are dimensionless,
the gauge field and the electromagnetic coupling constant $g$ possess mass
dimension $1$, and the Skyrme coupling constant $\lambda $ has mass
dimension $-1$.

It is important to emphasize that the system is at vacuum when $%
\Sigma(\chi)=1$, indicating that the field $\chi$ decouples from the gauge
field, reducing (\ref{3}) to the gauged BPS baby Skyrme model investigated
in Ref. \cite{Adam_12}.

The Euler-Lagrange equations obtained from the Lagrangian density (\ref{3})
are
\begin{equation}
\partial _{\mu }\left( \Sigma F^{\mu \nu }\right) =g^{2}j^{\nu }\text{,}
\label{xga}
\end{equation}%
\begin{equation}
D_{\mu }\vec{J}^{\mu }+\frac{\partial V}{\partial \phi _{n}}\hat{n}\times
\vec{\phi}=0\text{,}  \label{xg2}
\end{equation}%
\begin{equation}
\partial _{\mu }\partial ^{\mu }\chi +\frac{1}{4}\Sigma _{\chi }F_{\mu \nu
}F^{\mu \nu }+V_{\chi }=0\text{,}  \label{xg1}
\end{equation}%
where $\Sigma _{\chi }=\partial \Sigma /\partial \chi $, $V_{\chi}=\partial
V/\partial \chi $ and $j^{\mu }=\hat{n}\cdot \vec{J}^{\mu }$ is the
conserved current density with
\begin{equation}
\vec{J}^{\mu }=\lambda ^{2}[\vec{\phi}\cdot (D^{\mu }\vec{\phi}\times D^{\nu
}\vec{\phi})]D_{\nu }\vec{\phi}\text{.}  \label{jjkk}
\end{equation}

In the present {study, we are} interested in stationary soliton solutions of
the model described by the Lagrangian density (\ref{3}). In this sense, from
Eq. (\ref{xga}), we extract an important information via the stationary
Gauss law
\begin{equation}
\partial _{i}(\Sigma \partial _{i}{A_{0}})=g^{2}\lambda ^{2}A_{0}(\hat{n}%
\cdot \partial _{i}\vec{\phi})^{2}\text{,}  \label{Gauss1sp}
\end{equation}%
which is identically satisfied by gauge condition $A_{0}=0$, allowing us to
choose such a condition henceforth. Consequently, the system solutions only
carry on magnetic flux. Furthermore, from Eq. (\ref{xga}), we also obtain
the stationary Amp\`{e}re law, which is given by,
\begin{equation}
\partial _{i}\left( \Sigma B\right) +g^{2}\lambda ^{2}(\hat{n}\cdot \partial
_{i}\vec{\phi})Q=0\text{,}  \label{Amperestat}
\end{equation}%
where $B=F_{12}=\epsilon _{ij}\partial _{i}A_{j}$ is the magnetic field and $%
Q$ is
\begin{equation}
Q=\vec{\phi}\cdot (D_{1}\vec{\phi}\times D_{2}\vec{\phi})=q+\epsilon
_{ij}A_{i}(\hat{n}\cdot \partial _{j}\vec{\phi})\text{.}  \label{qqs}
\end{equation}%
The quantity $q$ related to the topological charge density of the Skyrme
field is given by
\begin{equation}
q=\frac{1}{2}\epsilon _{ij}\vec{\phi}\cdot (\partial _{i}\vec{\phi}\times
\partial _{j}\vec{\phi}),  \label{qqds}
\end{equation}%
such that the topological charge or topological degree (or winding number)
of the Skyrme field reads as
\begin{equation}
\deg [\vec{\phi}]=-\frac{1}{4\pi }\int q\,d^{2}\mathbf{x}=k,  \label{degph}
\end{equation}%
where $k\in \mathds{Z} \setminus 0$.

The stationary equation for the Skyrme field obtained from Eq. (\ref{xg2})
reads
\begin{equation}
D_{k}\vec{J}_{k}-\frac{\partial V}{\partial \phi _{n}}\hat{n}\times \vec{\phi%
}=0\text{,}  \label{phistat}
\end{equation}%
whereas, from Eq. (\ref{xg1}), we obtain for the $\chi $-field
\begin{equation}
\partial _{k}\partial _{k}\chi -\frac{1}{2}\Sigma _{\chi }B^{2}-V_{\chi }=0.
\label{xistat}
\end{equation}

In the next section, we implement the Bogomol'nyi procedure \cite{Bogomol}
to investigate the conditions under which the model (\ref{3}) engenders
self-dual or BPS configurations that minimize the energy of the system.

\subsection{The BPS structure}

The starting point is the stationary energy density of the model defined by
the Lagrangian density (\ref{3}), {which reads as}
\begin{equation}
\varepsilon=\frac{1}{2g^{2}}\Sigma B^{2}+\frac{\lambda ^{2}}{2}Q^{2}+\frac{1%
}{2}\left( \partial _{k}\chi \right) ^{2}+V\left( \phi _{n},\chi \right) ,
\label{13}
\end{equation}%
where we have used the gauge condition $A_{0}=0$. In order to ensure a
finite energy, the following boundary conditions must be satisfied:
\begin{eqnarray}
\lim_{\left\vert \mathbf{x}\right\vert \rightarrow \infty }\sqrt{\Sigma }%
B=0,~\ \lim_{\left\vert \mathbf{x}\right\vert \rightarrow \infty }Q=0,~\
\lim_{\left\vert \mathbf{x}\right\vert \rightarrow \infty }\partial
_{k}\chi=0,~\ \lim_{\left\vert \mathbf{x}\right\vert \rightarrow \infty
}V\left( \phi _{n},\chi \right) =0.  \label{eng2}
\end{eqnarray}

In what follow, the total de energy reads
\begin{equation}
E=\int \left[ \frac{1}{2g^{2}}\Sigma B^{2}+\frac{\lambda ^{2}}{2}Q^{2}+\frac{%
1}{2}\left( \partial _{k}\chi \right) ^{2}+V\right] d^{2}\mathbf{x},
\label{14a}
\end{equation}%
and to implement the Bogomol'nyi {\ procedure, we} introduce three auxiliary
functions, $W\equiv W(\phi _{n})$, $Z\equiv Z(\phi _{n})$ and the functions $%
F_{i}\equiv F_{i}(\chi )$ (with $i=1,2$), that allows us to express total
energy as 
\begin{align}
E& =\int \left[ \frac{\left( \Sigma B\pm g^{2}\lambda ^{2}W\right) ^{2}}{%
2g^{2}\Sigma }+\frac{1}{2}\left( \partial _{k}\chi \mp \epsilon _{kj}\frac{%
\partial F_{j}}{\partial \chi }\right) ^{2}+\frac{\lambda ^{2}}{2}\left(
Q\mp Z\right) ^{2} \right.   \notag \\[0.2cm]
&\hspace{1cm} \left.\mp \lambda ^{2}BW\pm \lambda ^{2}QZ\pm \epsilon
_{kj}\partial _{k}F_{j} +V-\frac{g^{2}\lambda ^{4}}{2}\frac{W^{2}}{\Sigma }-\frac{\lambda
^{2}}{2}Z^{2}-\frac{1}{2}\left( \frac{\partial F_{j}}{\partial \chi }\right)
^{2}\right] d^{2}\mathbf{x}.  \label{14}
\end{align}

Now, by setting
\begin{equation}
Z=\frac{\partial W}{\partial \phi _{n}},
\end{equation}%
and using Eq. (\ref{qqs}), the expression $\mp \lambda ^{2}BW\pm \lambda
^{2}QZ$ becomes%
\begin{equation}
\pm \lambda ^{2}\epsilon _{ij}\partial _{j}\left( A_{i}W\right) \pm \lambda
^{2}q\left( \frac{\partial W}{\partial \phi _{n}}\right) ,
\end{equation}%
with $q$ defined in Eq. (\ref{qqds}).  Further, by setting
\begin{equation}
V=\frac{g^{2}\lambda ^{4}}{2}\frac{W^{2}}{\Sigma }+\frac{\lambda ^{2}}{2}%
\left( \frac{\partial W}{\partial \phi _{n}}\right) ^{2}+\frac{1}{2}\left(
\frac{\partial F_{j}}{\partial \chi }\right) ^{2},  \label{bpspotential}
\end{equation}%
we establish a relation between the potential and the respective superpotentials. Next, the vacuum condition in Eq. (\ref{eng2}), allows us to establish
\begin{equation}
\lim_{\left\vert \mathbf{x}\right\vert \rightarrow \infty }W=0,~\
\lim_{\left\vert \mathbf{x}\right\vert \rightarrow \infty }\frac{\partial W}{
\partial \phi _{n}}=0,~\ \lim_{\left\vert \mathbf{x}\right\vert \rightarrow
\infty }\frac{\partial F_{j}}{\partial \chi }=0,  \label{WWsbc}
\end{equation}%
by considering $\Sigma ^{-1}$ a well-behaved function. Here it is
interesting to point out that both $W(\phi_n)$ and $F_i(\chi)$ play the role
of superpotential functions, the first one for the Skyrme field $\phi$ and
the second for the scalar field $\chi$. The choosing of both the
superpotentials must consider that the resultant self-dual potential (\ref%
{bpspotential}) ensures the finite-energy requirement, i.e., the energy
density (\ref{13}) must be null when the fields attain their vacuum values.

The first condition in (\ref{WWsbc}) tell us the integral of the total
derivative $\epsilon _{ij}\partial_{j}\left( A_{i}W\right) $ is zero, i.e.,
\begin{equation}
\int \epsilon _{ij}\partial _{j}\left( A_{i}W\right) ~d^{2}\mathbf{x}=0.
\end{equation}
This way, the considerations so far allow us to write Eq. (\ref{14}) as
\begin{eqnarray}
E=E_{_{\text{BPS}}}+\frac{1}{2g^{2}}\int \frac{\left( \Sigma B\pm
g^{2}\lambda ^{2}W\right) ^{2}}{\Sigma }d^{2}\mathbf{x}+\frac{\lambda ^{2}}{2}\int \left( Q\mp \frac{\partial W}{\partial \phi
_{n}}\right) ^{2}d^{2}\mathbf{x}+\frac{1}{2}\int \left( \partial _{k}\chi \mp \epsilon _{kj}\frac{\partial
F_{j}}{\partial \chi }\right) ^{2}d^{2}\mathbf{x},
\end{eqnarray}%
with $E_{_{\text{BPS}}}$ defined by
\begin{equation}
E_{_{\text{BPS}}}=\int \left[ \pm \lambda ^{2}q\left( \frac{\partial W}{%
\partial \phi _{n}}\right) \pm \epsilon _{kj}\left( \partial
_{k}F_{j}\right) \right] d^{2}\mathbf{x}\ >0,  \label{denBPS}
\end{equation}%
where the first term is the Skyrmion contribution to the BPS energy and the
second stands for the additional contribution coming from the $\chi $-field
sector.

The total energy satisfy the inequality $E\geq E_{_{\text{BPS}}}$, attaining
its minimum $E=E_{_{\text{BPS}}}$ when the Bogomol'nyi
bound is saturated, i.e., the fields satisfying the following set of
equations:
\begin{eqnarray}
\Sigma B &=&\mp g^{2}\lambda ^{2}W,  \label{bpsx1} \\[0.2cm]
Q &=&\pm \frac{\partial W}{\partial \phi _{n}},  \label{bpsx2} \\[0.2cm]
\partial _{k}\chi &=&\pm \epsilon _{kj}\frac{\partial F_{j}}{\partial \chi }.
\label{bpsx3}
\end{eqnarray}%
The set above defines the self-dual or BPS equations of the model (\ref{3}). The solutions of these BPS equations can also be considered classical solutions belonging to an extended supersymmetric model \cite{witten, spector} whose bosonic sector would be given by the Lagrangian density (\ref{3}). Some studies concerning skyrmions in the SUSY field theory context, including compactons, can be found, for example, in Refs. \cite{r6,Queiruga0, Sasaki01} for nongauged models and in Ref. \cite{Sasaki02} for the gauged ones.

Lastly, we highlight that the presence of the neutral scalar field $\chi $
in our model (\ref{3}) demands the insertion of the superpotentials $%
F_i(\chi)$ to complete its BPS structure. Below, the superpotentials $%
F_i(\chi)$ will be chosen to generate suitable solutions in the scalar
sector to define the magnetic permeability $\Sigma(\chi)$.

\subsection{BPS Skyrmions with radial symmetry}

We now consider rotationally symmetric solitons and also we set, without
loss of generality, $\hat{n}=(0,0,1)$ such that $\phi_{n}=\phi _{3}$. The
standard ansatz for the Skyrme field is
\begin{equation}
\vec{\phi}\left( r,\theta \right) =\left(
\begin{array}{c}
\sin f(r)\cos (N\theta ) \\
\sin f(r)\sin (N\theta ) \\
\cos f(r)%
\end{array}%
\right) \text{,}  \label{5}
\end{equation}%
where $r$ and $\theta $ are polar coordinates, $N=\deg [\vec{\phi}]$ is the
winding number presented in Eq. (\ref{degph}), and $f(r)$ is a well behaved
function obeying the boundary conditions%
\begin{equation}
f(0)=\pi \,\text{,}\;\;\lim_{r\rightarrow \infty }f(r)=0\text{.}
\end{equation}%
By convenience, we introduce the following field redefinition \cite{Adam_12}:%
\begin{equation}
\cos f(r)=1-2h(r)\text{,}
\end{equation}%
with the field $h(r)$ obeying%
\begin{equation}
h(0)=1\,\text{,}\;\;\lim_{r\rightarrow \infty }h(r)=0\text{.}  \label{5a}
\end{equation}%
Further, for the gauge field components and neutral scalar field we assume
\begin{equation}
A_{i}=-\epsilon _{ij}x_{j}\frac{Na(r)}{r^{2}}\text{,}\;\;\chi =\chi (r),
\label{6}
\end{equation}%
respectively. The fields $a(r)$ and $\chi(r)$ are regular functions
satisfying, at the origin, the conditions
\begin{equation}
a(0)=0\,\text{,}\;\;\chi (0)=\chi _{0}\text{,}  \label{6a}
\end{equation}%
whereas for the asymptotic limit we require%
\begin{equation}
\;\lim_{r\rightarrow \infty }a(r)=a_{\infty }\,\text{,}\;\;\;\lim_{r%
\rightarrow \infty }\chi (r)=\chi _{\infty }\text{,}  \label{6b}
\end{equation}%
where $\chi _{0}$, $\chi _{\infty }$ and $a_{\infty }$ are finite constants.

The magnetic flux is easily calculated to be
\begin{equation}
\Phi =2\pi \int_{0}^{R}Brdr=2\pi Na(R)=2\pi Na_{R}\text{,}  \label{9}
\end{equation}%
where $a_{R}=a(R)$ is a real constant, whereas $R>0$ defines the maximum
size of the configuration characterizing the type of solution, i.e., $R$ is
finite for compact solutions and infinite for noncompact solutions (see,
e.g., Refs. \cite{Adam_12, Casana_16, Casana_17}). As well as their
counterpart model \cite{Adam_12}, the magnetic flux (\ref{9}) is
nonquantized since $a_{R}$ and $a_{\infty }$ (when $R\rightarrow \infty $)
belong to the interval $\left\langle -1,0\right] $. It is a characteristic
property of Skyrme models that for sufficiently large values of the
electromagnetic coupling $g$, the quantities $a_{R}$ or $a_{\infty }$ tend
to $-1$ and, in such a limit, the magnetic flux becomes ``quantized" in
units of $2\pi $.

%%%%%%%%%%%%%%%%%%%%%%%%%

For the radial superpotentials, we take
\begin{equation}
W\equiv W(h(r)),  \label{Fi0}
\end{equation}
and
\begin{equation}
F_{i}=-\epsilon _{ij}x_{j}\frac{\mathcal{W}(\chi(r) )}{r^{2}}.  \label{Fi}
\end{equation}
For convenience, we write the radial version of the BPS potential (\ref%
{bpspotential}) as the sum of two contributions
\begin{equation}
V=V^{(\Sigma)}+V^{(\chi)},  \label{VsigXi}
\end{equation}
where
\begin{equation}
V^{(\Sigma)}=\frac{g^{2}\lambda ^{4}}{2}\frac{W^{2}}{\Sigma }+\frac{\lambda
^{2}}{8}W_{h}^{2}, \;\; V^{(\chi)}=\frac{1}{2r^{2}}\mathcal{W}_{\chi }^{2}%
\text{,}  \label{16}
\end{equation}
being $W_{h}=\partial W/\partial h$ and $\mathcal{W}_{\chi } =\partial
\mathcal{W}/\partial \chi $. The vacuum condition for the potential $V(h,
\chi)$ provides the boundary conditions for $W$ and $W_{h}$ as set in Eq. (%
\ref{WWsbc}), whereas for the $\mathcal{W}_{\chi}$ we have
\begin{equation}
\lim_{r\rightarrow \infty }\mathcal{W}_{\chi }=0.
\end{equation}
Further, the superpotentials $W(h)$ and $\mathcal{W}(\chi)$ are regular
functions obeying the following boundary conditions at the origin:
\begin{equation}
W\left( h(0)\right) =W\left( 1\right) =W_{0}\text{,}  \label{bc1}
\end{equation}%
\begin{equation}
\mathcal{W}\left( \chi (0)\right) =\mathcal{W}\left( \chi _{0}\right) =%
\mathcal{W}_{0}\text{,}  \label{bc2}
\end{equation}%
{whereas for the asymptotic limit we get}
\begin{equation}
\lim_{r\rightarrow \infty }W(h)=W(0)=0\text{,}  \label{bc3}
\end{equation}%
\begin{equation}
\lim_{r\rightarrow \infty }\mathcal{W}\left( \chi \right) =\mathcal{W}\left(
\chi _{\infty }\right) =\mathcal{W}_{\infty }\text{,}  \label{bc4}
\end{equation}%
where $W_{0}$, $\mathcal{W}_{0}$ and $\mathcal{W}_{\infty }$ are constants, {%
and they are in accordance} with the boundary conditions (\ref{5a}), (\ref%
{6a}) and (\ref{6b}).

We observe in the BPS potential (\ref{VsigXi}) the contribution $V^{(\chi)}$%
, which, per the proposal in (\ref{Fi}), becomes out to be an explicit
function of the radial coordinate. Nevertheless, we emphasize that such
explicit dependence in the radial coordinate allows circumventing {\cite%
{Bazeia_13}} the Derrick-Hobard scaling theorem \cite{Hobart_10}%
. Consequently, it allows us to attain stable kinklike solutions in the $%
\chi $-field sector, as previously considered in literature in different
scenarios \cite{Bazeia_780,Bazeia_17, Bazeia_105024,
Andrade_056014,Casana_056014}.

Let us now show the BPS energy density under radial symmetry which, as well
as BPS potential (\ref{VsigXi}), is interesting to be expressed as the sum
of two contributions
\begin{equation}
\varepsilon_{_{\text{BPS}}}=\varepsilon_{\Sigma }+\varepsilon_{\chi }\text{,}
\label{22B}
\end{equation}%
where we have defined
\begin{equation}
\varepsilon _{\Sigma }=\frac{g^{2}\lambda ^{4}}{\Sigma }W^{2}+\frac{\lambda
^{2}}{4}{W}_{h}^{2}\!\!\quad \text{and}\quad \varepsilon _{\chi }=\frac{
\mathcal{W}_{\chi}^{2}}{r^{2}}\!\!\text{ ,}  \label{23B0}
\end{equation}
respectively. The quantity $\varepsilon_{\Sigma}$ represents the energy
density associated purely with the new skyrmion {configurations, while $%
\varepsilon_{\chi}$ is the contribution associated with the kinklike
solution engendered} by scalar field $\chi$.

The energy's lower bound (or Bogomol'nyi bound) in Eq. (\ref{denBPS})
becomes
\begin{eqnarray}
E_{_{\text{BPS}}} &=& \pm 2{\pi }\left( \lambda ^{2}NW_{0}+\Delta \mathcal{W}%
\right) > 0\text{,}  \label{13c}
\end{eqnarray}
where the integration has been performed using the boundary conditions (\ref%
{bc1}), (\ref{bc2}), (\ref{bc3}), and (\ref{bc4}). Furthermore, we have
defined the quantity $\Delta\mathcal{W=W}_{\infty}-\mathcal{W}_{0}$, such
that the upper (lower) sign in (\ref{13c}) describes the self-dual solitons
(antisolitons) corresponding to $N$ and $\Delta \mathcal{W}$ positive
(negative) quantities.

The set of BPS equations (\ref{bpsx1}), (\ref{bpsx2}) and (\ref{bpsx3}),
reads
\begin{equation}
B=\frac{N}{r}\frac{da}{dr}=\mp {g^{2}\lambda ^{2}}\frac{W}{\Sigma },
\label{21}
\end{equation}%
\begin{equation}
Q=\frac{2N}{r}(1+a)\frac{dh}{dr}=\mp \frac{W_{h}}{2},  \label{22}
\end{equation}%
\begin{equation}
\frac{d\chi }{dr}=\pm \frac{\mathcal{W}_{\chi }}{r},  \label{23}
\end{equation}%
respectively. We now observe that the third BPS equation (\ref{23}) has no
explicit dependence on the functions $h(r)$ and $a(r)$, allowing it to be
solved separately by adequately choosing the superpotential $\mathcal{W}
(\chi)$. The kinklike solution obtained from Eq.(\ref{23}) defines the
magnetic permeability $\Sigma(\chi) $ appearing explicitly in the BPS
equation (\ref{21}). Consequently, once $\Sigma(\chi) $ is known, we proceed
to solve the two self-dual equations (\ref{21}) and (\ref{22}).

At this level, to define the model entirely, we observe that it is necessary
to choose the functions $W(h)$ and $\mathcal{W}(\chi)$. Thus, for our
analysis, we may consider functions $W(h)$ already studied in other BPS
Skyrme models (see, e.g., Refs. \cite{Adam_12, Casana_16, Casana_17,
Casana_18}). For the function $\mathcal{W}(\chi)$ associated with the
neutral scalar field $\chi $, we will consider the ones engendering kinklike
solutions. Furthermore, such as in other Skyrme's models \cite{Adam_12,
Casana_16, Casana_17}, we have verified that the our BPS model here studied
support compact and noncompact skyrmions. Namely, by taking a given
superpotential $W(h)$, the type of skyrmion engendered is determined by
analyzing how the fields approach their respective vacuum values. For such a
purpose, it is considered a superpotential $W(h)$ whose behavior is
\begin{equation}
W(h) \approx W_{R}^{\left( \sigma \right) }h^{\sigma }\text{,}  \label{Wh}
\end{equation}%
near the vacuum, where $W_{R}^{\left( \sigma \right) }>0$ and $\sigma >1$.
Thus, the analysis is performed by taking now the boundary conditions
\begin{equation}
h(r) =0\text{, \ }a(r) =a_{R}\text{, \ \ }W\left( h(r) \right) =0\text{,}
\end{equation}
being $R$ and $a_{R}$ are quantities already introduced in (\ref{9}). The $%
\sigma $ values characterize the soliton solutions. Indeed, for $1<\sigma <2$%
, we have the soliton called compacton, whose vacuum value is reached in a
finite radius $R$ (the compacton's radius) and remains in the vacuum for all
$r>R$, see e.g. Refs. \cite{Casana_16, Casana_17}. Otherwise, for $%
\sigma\geq 2$ we have the radius $R\rightarrow \infty$ which provides
extended or noncompact configurations that result in localized or
delocalized solitons, see e.g. Refs. \cite{Casana_16, Casana_17, Casana_18}.

In the remainder of the manuscript, we only study solitonic solutions by
adopting the upper sign, i.e., $N>0$ and $\Delta\mathcal{W}>0$.
Specifically, our principal focus is to analyze compact skyrmion's new
properties that could appear when immersed in a magnetic medium.

\section{BPS skyrmions in magnetic media: compactons\label{SecVI}}

To investigate BPS compactons, we choose the superpotential as being
\begin{equation}
W(h)=W_{0}h^{3/2}\text{,}  \label{superWh}
\end{equation}
where $W_{0}$ is a constant. An important detail is that this superpotential
provides a potential $V^{(\Sigma)}\propto h$ when $r\rightarrow R$, this
being analogous to the so-called ``old baby Skyrme potential'' \cite{Adam_12}%
.

The next step is to specify some functional form for both the superpotential
$\mathcal{W}(\chi)$ and the magnetic permeability $\Sigma(\chi)$. For such a
goal, we will consider two superpotentials (that engendering the $\chi^{4}$
and $\chi^{6}$ models, respectively) supplying kinklike solitons that we use
to define the magnetic permeability appearing in the BPS equation (\ref{21}%
). This way, it will be possible to analyze how the magnetic medium modifies
the shape of the compact BPS skyrmions.

\subsection{$\protect\chi^4$ medium}

{In our first scenario, we consider a superpotential $\mathcal{W}(\chi)$
engendering a $\chi^4$ model, i.e.,}
\begin{equation}
\mathcal{W}(\chi )=\alpha \chi -\frac{\alpha }{3}\chi ^{3},  \label{spotchi2}
\end{equation}
where $\alpha $ is a nonnegative real parameter. In the literature, we find
this superpotential applied in different contexts, for example, in the study
of vortices with internal structures \cite{Bazeia_780, Andrade_056014,
Bazeia_17, Casana_056014}, magnetic monopoles \cite{Bazeia_105024},
skyrmion-like configurations \cite{Bazeia_1947, Bazeia_423}, and the
behavior of massless Dirac fermions in a skyrmion-like background \cite%
{Bazeia_779}. Then, by using (\ref{spotchi2}) in (\ref{23}) and solving the
BPS equation, we obtain the exact kinklike solution
\begin{equation}
\chi (r)=\frac{r^{2\alpha }-r_{0}^{2\alpha }}{r^{2\alpha }+r_{0}^{2\alpha }},
\end{equation}%
with $r_{0}$ being an arbitrary positive constant such that $\chi(r_{0})=0$.
This solution satisfies the boundary conditions $\chi (0)=\chi _{0}=-1$ and $%
\chi (\infty)=\chi _{\infty }=1$. In this case, the BPS bound (\ref{13c})
for the energy becomes
\begin{equation}
E_{_{\text{BPS}}}=2\pi \lambda ^{2}NW_{0}+\frac{8}{3}\alpha \pi \text{,}
\end{equation}%
where the last term is the contribution from the neutral scalar field $\chi$.

We must now select the function defining the magnetic permeability of the
medium where the compact skyrmions will stay immersed. For this, we have
inspired ourselves with investigations regarding the pattern formation of
ringlike solitons \cite{Wang,Dai} in a two-dimensional Bose-Einstein
condensate interacting with a harmonic oscillator, which provides periodic
modulation to the system. Our approach will adjust the magnetic permeability
using a periodic function to modulate the topological structures. This way,
we set it as
\begin{equation}
\Sigma (\chi )=\frac{1}{\sin ^{2}(m\pi \chi )}\text{,}  \label{df1}
\end{equation}
where $m\in\mathds{N}$. Further, functions like that also has been used, for
example, in the study of kinklike solutions \cite{Bazeia_383} and vortexlike
structures \cite{Bazeia_17, Casana_056014}.

This way, the energy density $\varepsilon _{\Sigma }$ results be
\begin{equation}
\varepsilon _{\Sigma }=g^{2}\lambda ^{4}W_{0}^{2}h^{3}\sin ^{2}(m\pi \chi )+%
\frac{9}{16}\lambda ^{2}W_{0}^{2}h\text{.}
\end{equation}%
Already the BPS equations (\ref{21}) and (\ref{22}) to be investigated
assume the form
\begin{equation}
\frac{N}{r}(1+a)\frac{dh}{dr}+\frac{3}{8}W_{0}h^{1/2}=0,  \label{bpsC11}
\end{equation}
\begin{equation}
\frac{N}{r}\frac{da}{dr}+{g^{2}\lambda ^{2}W}_{0}\sin ^{2}\left( m\pi \frac{%
r^{2\alpha }-r_{0}^{2\alpha }}{r^{2\alpha }+r_{0}^{2\alpha }}\right)
h^{3/2}=0,  \label{bpsC1}
\end{equation}
{which we must solve with suitable field boundary conditions to provide
compact skyrmions in the current scenario.}

{Before we solve the BPS equations numerically, let us show the behavior of
the field profiles close to the boundary values. Thus, near to the origin,
by taking the more relevant terms, the Skyrme and gauge field profiles
behave as,}
\begin{eqnarray}
h(r) &=&1-\frac{3}{2^{4}}\frac{W_{0}}{N}r^{2}+\frac{3^{2}}{2^{10}}\frac{%
W_{0}^{2}}{N^{2}}r^{4}-\frac{3}{2^{4}}\frac{C_{0}W_{0}}{\left( \alpha +1\right) \left( 2\alpha
+1\right) N^{2}}\frac{r^{4\alpha +4}}{r_{0}^{4\alpha }}+\cdots \text{,}
\end{eqnarray}%
\begin{eqnarray}
a(r) &=&-\frac{2C_{0}}{\left( 2\alpha +1\right) N}\frac{r^{4\alpha +2}}{%
r_{0}^{4\alpha }}+\frac{3^{2}}{2^{5}}\frac{C_{0}W_{0}}{\left( \alpha
+1\right) N^{2}}\frac{r^{4\alpha +4}}{r_{0}^{4\alpha }}+\frac{4C_{0}}{\left( 3\alpha +1\right) N}\frac{r^{6\alpha +2}}{%
r_{0}^{6\alpha }}+\cdots \text{,}
\end{eqnarray}%
where we have defined the constant $C_{0}=\pi ^{2}m^{2}g^{2}\lambda
^{2}W_{0} $. {In the sequel, we present the behavior, close to the origin,
of both the magnetic field and the energy density $\varepsilon _{\Sigma}(r) $%
. Then, the magnetic field reads}
\begin{eqnarray}
B(r) &=&-4C_{0}\frac{r^{4\alpha }}{r_{0}^{4\alpha }}+\frac{3^{2}}{2^{3}}%
\frac{C_{0}W_{0}}{N}\frac{r^{4\alpha +2}}{r_{0}^{4\alpha }}-\frac{3^{3}}{2^{8}}\frac{C_{0}W_{0}^{2}}{N^{3}}\frac{r^{4\alpha +4}}{%
r_{0}^{4\alpha }}+8C_{0}\frac{r^{6\alpha }}{r_{0}^{6\alpha }}+\cdots \text{,}
\label{B01}
\end{eqnarray}%
{showing that it is null in $r=0$, whereas the energy density $\varepsilon
_{\Sigma}(r)$ reading as}
\begin{eqnarray}
\varepsilon _{\Sigma } &=&\frac{3^{2}}{2^{4}}\frac{\lambda ^{2}W_{0}^{2}}{N}-%
\frac{3^{3}}{2^{8}}\frac{\lambda ^{2}W_{0}^{3}}{N}r^{2}+\frac{3^{4}}{2^{14}}%
\frac{\lambda ^{2}W_{0}^{4}}{N^{2}}r^{4}+4C_{0}\lambda ^{2}W_{0}\frac{r^{4\alpha }}{r_{0}^{4\alpha }}+\cdots \text{%
,}  \label{E01}
\end{eqnarray}%
{presents a nonnull value at the origin.}

{On the other hand, near the vacuum values, i.e., when $r\rightarrow R$, the
field profiles} possess the following behavior:
\begin{equation}
h(r) \approx \frac{3^{2}}{2^{8}}\frac{R^{2}W_{0}^{2}}{\left( 1+a_{R}\right)
^{2}N^{2}}\left( r-R\right) ^{2}\text{,}
\end{equation}%
\begin{equation}
a(r) \approx a_{R}+\frac{1}{2^{2}}\frac{R^{2}}{N^{2}}C_{R}\left( r-R\right)
^{4}\text{,}
\end{equation}%
with the constant $C_{R}$ defined as%
\begin{equation}
C_{R}=\frac{3^{3}}{2^{12}}\frac{g^{2}\lambda ^{2}W_{0}^{4}R^{2}}{\left(
1+a_{R}\right) ^{3}N^{2}}\sin ^{2}\left( m\pi \frac{R^{2\alpha
}-r_{0}^{2\alpha }}{R^{2\alpha }-r_{0}^{2\alpha }}\right) \text{,}
\end{equation}
which depends on the new parameters $m$, $\alpha$ and $r_{0}$ that control the medium. Similarly, the first relevant terms of the magnetic field and energy density $\varepsilon_{\Sigma}$, when $r\rightarrow R$, are given by
\begin{equation}
B(r) =-\frac{R}{N}C_{R}\left( r-R\right) ^{3}+\cdots \text{,}
\end{equation}%
and%
\begin{eqnarray}
\varepsilon _{\Sigma } &=&\frac{3^{4}}{2^{12}}\frac{\lambda
^{2}W_{0}^{4}R^{2}}{\left( 1+a_{R}\right) ^{2}N^{2}}\left( r-R\right)
^{2}+\cdots\notag\\[0.2cm]
& &+\frac{3^{3}}{2^{12}}\frac{\lambda ^{2}W_{0}^{4}R^{4}}{\left(
1+a_{R}\right) ^{3}N^{4}}C_{R}\left( r-R\right) ^{6}+\cdots \text{.}
\end{eqnarray}

{We next solve the BPS system, Eqs. (\ref{bpsC1}) and (\ref{bpsC11}),
numerically to investigate how the magnetic permeability (\ref{df1})
modifies the soliton profiles. For such a purpose, we fix $N=1$, $W_{0}=1$, $%
\lambda =1$, $g=1$, by considering distinct values of both $\alpha$ and $m$,
besides, $r_{0}$ will have a specified value in each case.} It is important
to point out that although the fixed parameters also could be explored
(e.g., by running different values), {we have used only $\alpha$ and $m$ to
control the profile shape of the fields. Both parameters are} enough to
exhibit the main features {of the compact skyrmions immersed in the medium
defined by (\ref{df1}).}

\begin{figure}[t]
	\includegraphics[width=5cm]{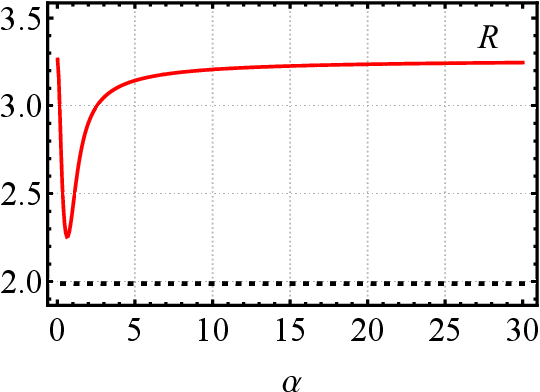} \hspace{0.5cm}%
	\includegraphics[width=5cm]{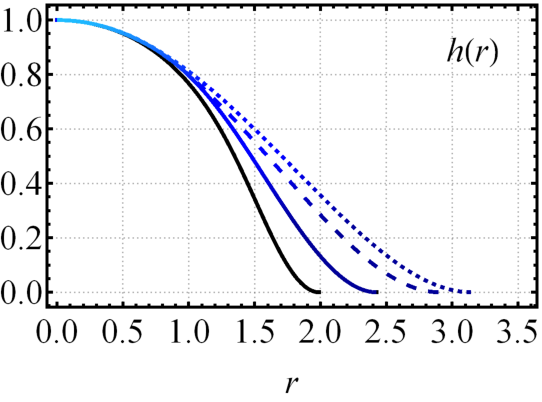}
	\caption{Depiction by assuming the magnetic permeability (\protect\ref{df1})
		with $r_{0}=0.5$, $m={1}$ and distinct values for $\protect\alpha$. Left:
		compacton radius $R$ vs. $\protect\alpha$ (solid red line) and the compacton
		radius of the standard case (black dot line). Right: the present Skyrme
		field (color lines) is depicted for $\protect\alpha=1$ (solid line), $%
		\protect\alpha=2$ (dashed line) and $\protect\alpha=5$ (dot line), and the
		solid black line represents the profile without the magnetic medium. }
	\label{Fig01}
\end{figure}

\begin{figure}[tbp]
	\includegraphics[width=5.2cm]{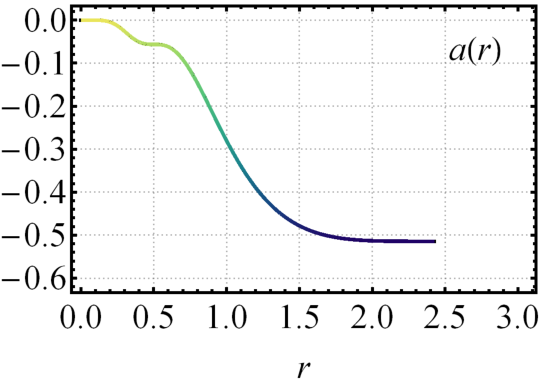} \hspace{0.5cm}%
	\includegraphics[width=5.cm]{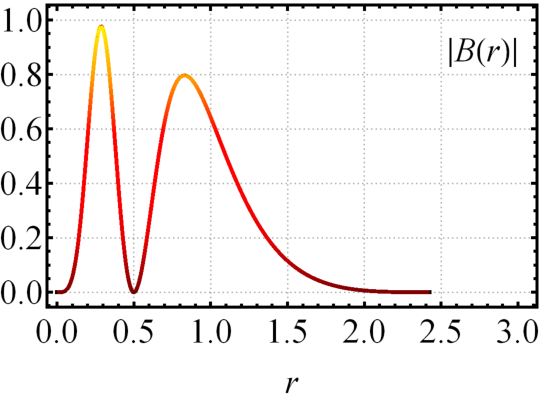} \hspace{0.5cm}%
	\includegraphics[width=5.cm]{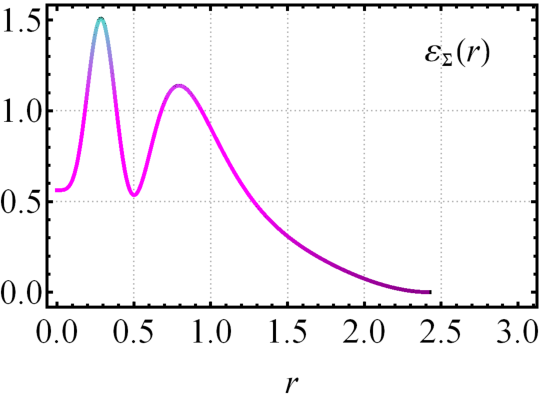}
	\includegraphics[width=5.3cm]{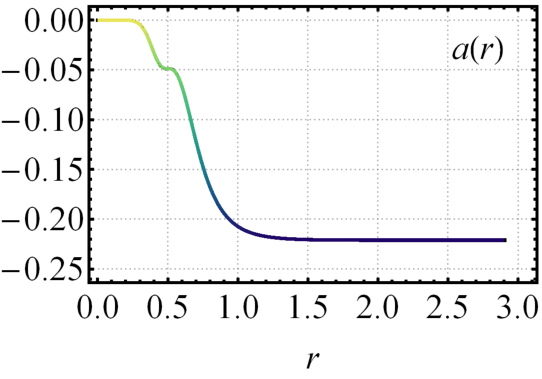} \hspace{0.5cm}%
	\includegraphics[width=5.cm]{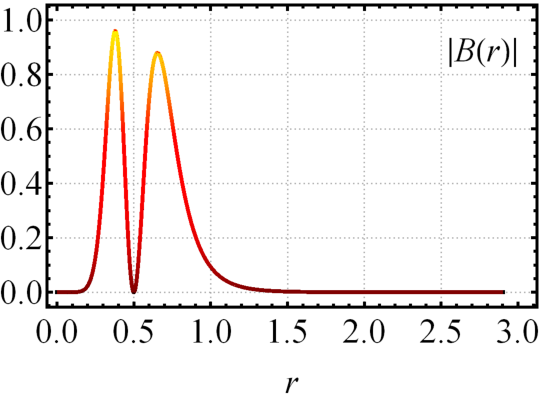} \hspace{0.5cm}%
	\includegraphics[width=5.cm]{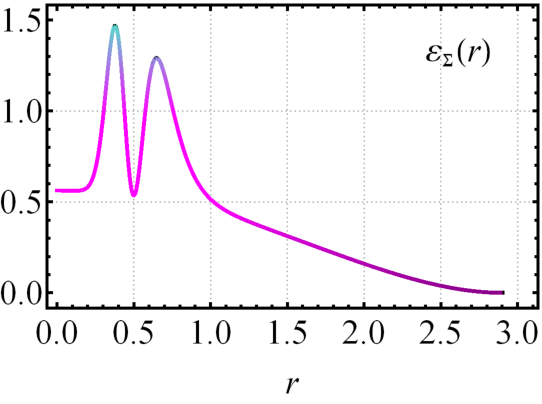}
	\caption{{The gauge field profiles} $a(r)$ ({left} panels),
		absolute value of the magnetic field $\vert B(r)\vert$ (middle panels) and	energy density $\protect\varepsilon_{\Sigma}(r)$ ({right}
		panels). We depict for $r_{0}=0.5$, $m=1$ with $\protect\alpha=1$ 
({top}) and $\protect\alpha=2$ ({bottom}).}
	\label{Fig02}
\end{figure}

\begin{figure}[tbp]
	\includegraphics[width=5.2cm]{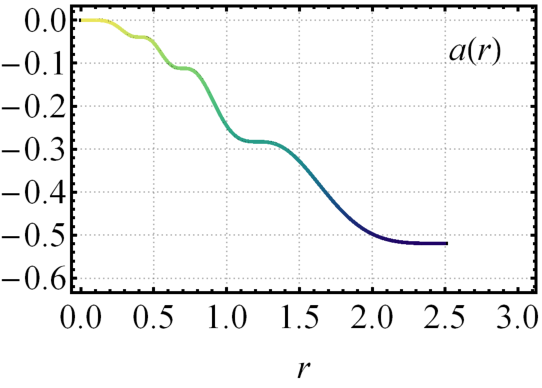} \hspace{0.5cm}%
	\includegraphics[width=5.cm]{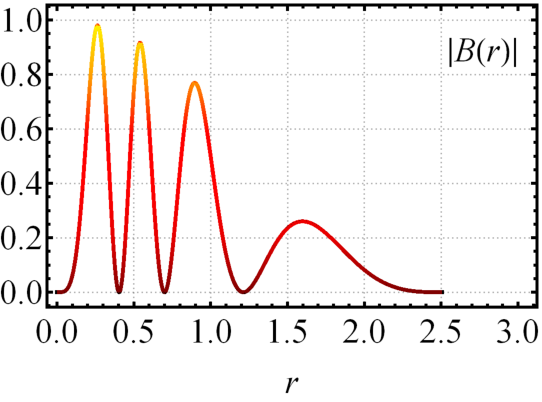} \hspace{0.5cm}%
	\includegraphics[width=5.cm]{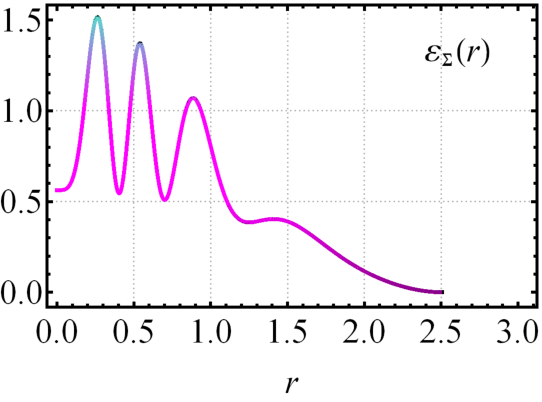}
	\includegraphics[width=5.3cm]{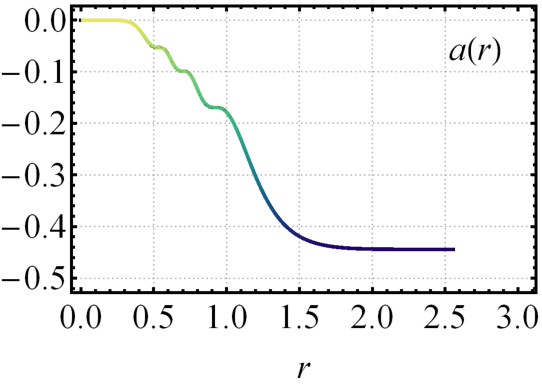} \hspace{0.5cm}%
	\includegraphics[width=5.cm]{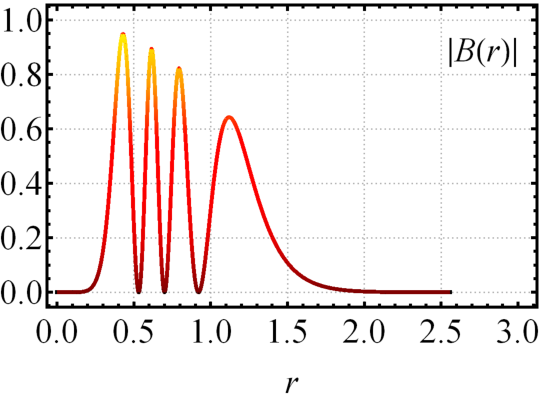} \hspace{0.5cm}%
	\includegraphics[width=5.cm]{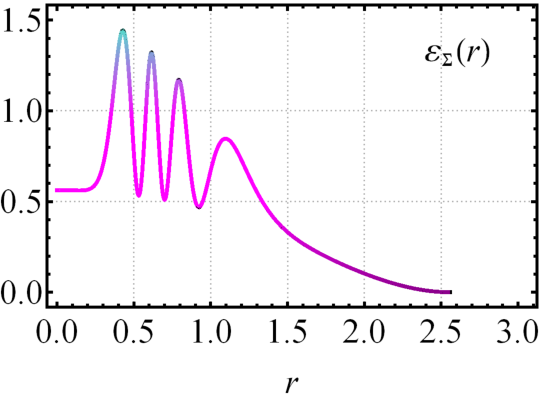}
	\caption{{The gauge field profiles} $a(r)$ (left panels),
		absolute value of the magnetic field $\vert B(r)\vert$ (middle panels) and energy density $\protect\varepsilon_{\Sigma}(r)$ (right panels). We depict for $r_{0}=0.7$, $m=2$ with $\protect\alpha=1$ (top) and $\protect\alpha=2$ (bottom).}
	\label{Fig04}
\end{figure}

\begin{figure*}[]
	\centering
	{\rule{0.\linewidth \includegraphics[width=6.2cm]{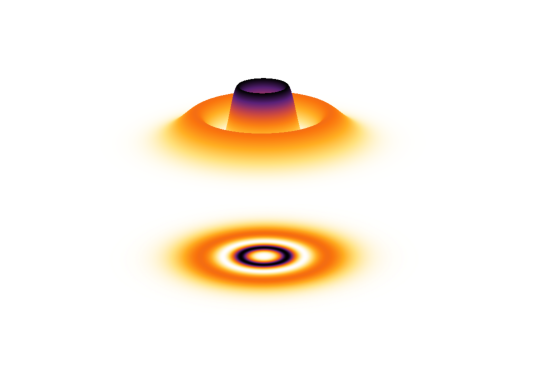}}{4cm}} {%
		\hspace{-0.6cm}\rule%
		{0.\linewidth\includegraphics[width=6.2cm]{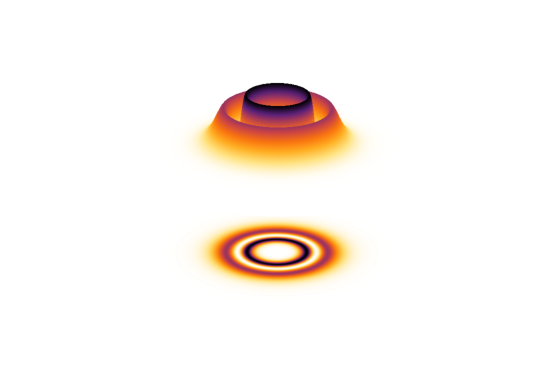}}{4cm}} {\hspace{-0.6cm%
		} \rule{0.\linewidth\includegraphics[width=6.2cm]{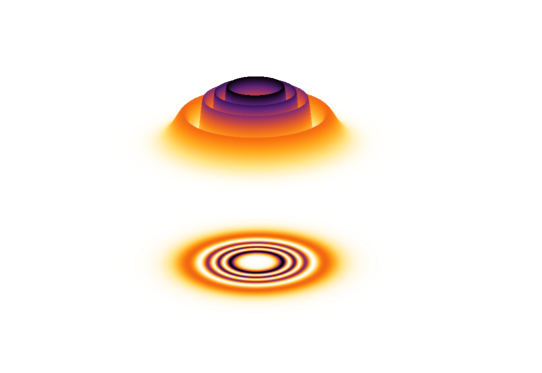}}{4cm}}\vspace{%
		-0.5cm}
	\caption{The magnetic field distributions (upper plots) with their
		corresponding projections in the plane (bottom plots). We depict for $%
		r_{0}=0.7$ with $m=1$ and $\protect\alpha=1$ (left plot), $m=1$ and $\protect%
		\alpha=2$ (middle plot), and $m=2$ and $\protect\alpha=2$ (right plot). }
	\label{Fig03}
\end{figure*}

{We have begun} our analysis by investigating how the compacton radius $R$ {%
changes as a function of the $\alpha$ parameter for a fixed $r_0$ and $m$,
shown in the left panel of Fig. \ref{Fig01}. Likewise, we have verified that
exists an interval $0<\alpha \leq \bar{\alpha}$ (here $\bar{\alpha}\simeq 0.6
$) that provides global} extreme values for the compacton radius. Namely,
for a sufficiently small $\alpha $ (close to zero) we {have the maximum}
compacton radius $R_{\max}\simeq 3.26$ which decreases until it {reaches the
minimum} value $R_{\min}\simeq 2.25$ when $\alpha= \bar{\alpha}$. From this
point on, i.e., for $\alpha >\bar{\alpha}$, the values of $R$ begins to
increase in a way that $R\rightarrow R_{\max}$. We also observe that the
magnetic medium modifies the compacton radius, which achieves higher values
than the ones acquired without the medium (we will call $R_{0} \simeq 1.99$,
see the horizontal black dotted line). Alike, in the right panel of Fig. \ref%
{Fig01}, we illustrate this effect on the Skyrme field profiles by depicting
$h(r)$ for some values of $\alpha> \bar{\alpha}$. So, we observe how the
parameter $\alpha$ modifies the profiles by showing the increase of the
compacton radius as $\alpha $ grows.

%%$$$$$$$$$$$$$$$$$$$

{Figure \ref{Fig02}} shows the profiles for the gauge field $a(r)$, magnetic
field $B(r)$ and energy density $\varepsilon _{\Sigma}(r)$, {all of them
well-behaved functions and in according to the respective behaviors
previously found at the boundary values. The gauge field $a(r)$ profiles
acquire a plateau format around $r_{0}$ that affect directly the profiles of
both the magnetic field and $\varepsilon _{\Sigma}(r)$, i.e., because of the
plateau, they acquire a ringlike shape centered at the origin. We also note
that when} $\alpha $ increases, the profiles become more localized around $%
r_{0}$. {On the other hand, the role played by the parameter $m$ is better
understood by examining Fig. \ref{Fig04}. We have verified that for a fixed $%
m$,} the gauge field engenders $\left(2 m-1\right) $-plateaus {that define,
in the profiles of both} $B(r)$ and $\varepsilon _{\Sigma}(r)$, an equal
number of {outer rings} around its center. {For a more precise} overview of
these features, we depict in Fig. \ref{Fig03} the effects induced in the
magnetic field profiles by the magnetic permeability (\ref{df1}) for
different values of $\alpha $ and $m$. It is notorious how $\alpha$ controls
the magnetic field distribution around the origin. That is, the size of the
inner region to the ringlike structures increases as $\alpha $ grows; that
effect arises because the rings agglomerate around $r_{0}$ when $\alpha$
growths (see left and middle plots in Fig. \ref{Fig03}). In addition, we see
the explicit formation of $2m-1$ outer rings, corresponding to the same
quantity of plateaus engendered in the gauge profile.

Therefore, our results show how the magnetic permeability (\ref{df1}),
induced by the kinklike soliton of the ${\chi^4}$-model, provides new
features to the profiles of both the magnetic field and the BPS energy
density of the compact skyrmions originally at the vacuum, i.e., when $%
\Sigma(\chi)=1$.

\subsection{$\protect\chi^{6}$ medium}

In this second scenario, we adopt a superpotential that engenders a $\chi^6$
model, so we consider
\begin{equation}
\mathcal{W}(\chi )=\frac{\alpha }{2}\chi ^{2}-\frac{\alpha }{4}\chi ^{4}%
\text{,}  \label{spotchi}
\end{equation}
{which has also been} used in the study of multilayered structures {\cite%
{Bazeia_17}. Then, by using (\ref{spotchi}) in the BPS equation (\ref{23}),
we obtain the kinklike solution generated by the $\chi^6$-model that is
given by }
\begin{equation}
\chi (r)=\frac{r^{\alpha }}{\sqrt{r^{2\alpha }+r_{0}^{2\alpha }}}\text{,}
\end{equation}%
{satisfying the following boundary values} $\chi _{0}=0$ and $\chi _{\infty
}=1$. {Likewise, now the BPS bound} (\ref{13c}) becomes
\begin{equation}
E_{_{\text{BPS}}}=2\pi \lambda ^{2}NW_{0}+\frac{\alpha \pi }{2}\text{,}
\end{equation}%
where the second {term is the} contribution from the neutral scalar field $%
\chi$.

To investigate the changes in the shape of the soliton originate from a $%
\chi^6$ model, we select the following magnetic permeability,
\begin{equation}
\Sigma (\chi )=\frac{1}{J_{0}^{2}(\gamma \chi )}\text{,}  \label{df2}
\end{equation}%
where $J_{0}$ is the zero-order Bessel function of the first kind and $%
\gamma \in\mathds{R}$. The primary motivation for this choice is the
ringlike patterns that solitons produce in Bessel photonic lattices when
interacting with a nonlinear medium, as mentioned in Refs. \cite{Kartashov1,
Kartashov2}. These studies specifically investigated the behavior of optical
radiation in a bulk cubic medium subjected to a transverse modulation of the
linear refractive index. Our current study will observe that the magnetic
permeability (\ref{df2}) plays a modulation role, contributing to the
raising of compact skyrmions with ring-shaped structures, too.

Taking $\Sigma(\chi)$ defined in (\ref{df2}), the BPS equations to be solved
are given by to the ones in (\ref{bpsC11}) and the second equation now reads
\begin{equation}
\frac{N}{r}\frac{da}{dr}+{g^{2}\lambda ^{2}W}_{0}J_{0}^{2}\left( \frac{%
\gamma r^{\alpha }}{\sqrt{r^{2\alpha }+r_{0}^{2\alpha }}}\right) h^{3/2}=0%
\text{.}  \label{bpscII}
\end{equation}
It is also important to write the energy density $\varepsilon_{\Sigma}(r)$
for the current scenario,
\begin{equation}
\varepsilon _{\Sigma }=g^{2}\lambda ^{4}W_{0}^{2}h^{3}J_{0}^{2}+\frac{9}{16}%
\lambda ^{2}W_{0}^{2}h\text{.}
\end{equation}

To continue, we write below the behavior of the field profiles near the
boundary values. Like this, around the origin, we obtain
\begin{eqnarray}
h(r) &=&1-\frac{3}{2^{4}}\frac{W_{0}}{N}r^{2}+\frac{3}{2^{10}}\frac{\left(
16g^{2}\lambda ^{2}-3\right) W_{0}^{2}}{N^{2}}r^{4}+\cdots\nonumber \\[0.3cm] &&+\frac{3}{2^{6}}\frac{\gamma ^{2}g^{2}\lambda ^{2}W_{0}^{2}}{\left( \alpha
+1\right) \left( \alpha +2\right) N^{2}}\frac{r^{2\alpha +4}}{r_{0}^{2\alpha
}}+\cdots \text{,} \\[0.2cm]
a(r) &=&-\frac{\lambda ^{2}g^{2}W_{0}}{2N}r^{2}+\frac{3^{2}}{2^{7}}\frac{%
g^{2}\lambda ^{2}W_{0}^{2}}{N^{2}}r^{4}+\cdots\nonumber\\[0.3cm]
&&+\frac{\gamma ^{2}g^{2}\lambda ^{2}W_{0}}{4N\left( 1+\alpha \right) }\frac{%
r^{2\alpha +2}}{r_{0}^{2\alpha }}+\cdots \text{.}
\end{eqnarray}

Besides, for the magnetic field and the energy density $\varepsilon
_{\Sigma }$ we get
\begin{eqnarray}
B(r) &\approx &-\lambda ^{2}g^{2}W_{0}+\frac{3^{2}}{2^{5}}\frac{g^{2}\lambda
^{2}W_{0}^{2}}{N}r^{2}+\cdots\notag\\[0.2cm]
&& +\frac{\gamma ^{2}g^{2}\lambda ^{2}W_{0}}{2}\frac{r^{2\alpha }}{%
r_{0}^{2\alpha }}+\cdots\text{,}  \label{bv2}
\end{eqnarray}
and
\begin{eqnarray}
\varepsilon _{\Sigma } &=&g^{2}\lambda ^{4}W_{0}^{2}+\frac{3^{2}}{2^{4}}%
\lambda ^{2}W_{0}^{2}-\frac{3^{2}}{2^{8}}\frac{\lambda ^{2}\left(16g^{2}\lambda ^{2}+3\right) W_{0}^{3}}{N}r^{2}+\cdots\notag\\[0.2cm]
&&-\frac{\gamma ^{2}g^{2}\lambda ^{4}W_{0}^{2}}{2}\frac{r^{2\alpha }}{r_{0}^{2\alpha }}+\cdots \text{,}
\end{eqnarray}
{respectively. Moreover, we observe that both are nonnull at the origin.}

\begin{figure}[t]
	\includegraphics[width=5.cm]{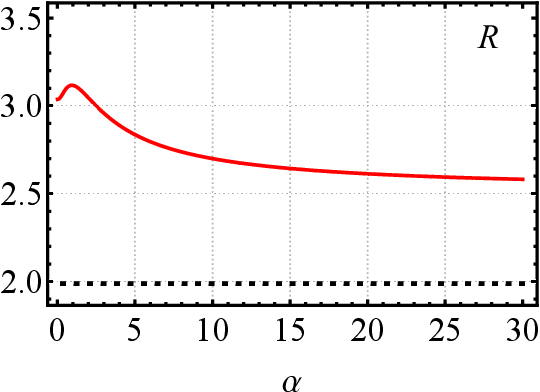}  \hspace{0.5cm}%
	\includegraphics[width=5.cm]{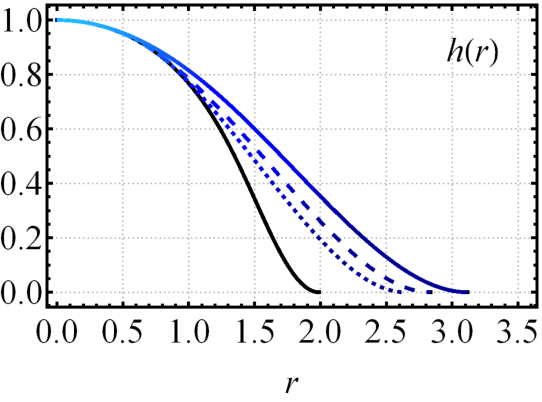}
	\caption{Depiction by assuming the magnetic permeability (\protect\ref{df2})
		with $\protect\gamma=6$ and distinct values for $\protect\alpha$. Left:
		compacton radius $R$ vs. $\protect\alpha$ (solid red line) and the compacton
		radius of the standard case (black dot line). Right: the present Skyrme
		field (color lines) is depicted for $\protect\alpha=1$ (solid line), $%
		\protect\alpha=5$ (dashed line), and $\protect\alpha= 15$ (dot line), and
		the solid black line represents the profile at the vacuum ($\Sigma(\protect%
		\chi)=1$).}
	\label{Fig1}
\end{figure}

\begin{figure}[tbp]
	\includegraphics[width=5.2cm]{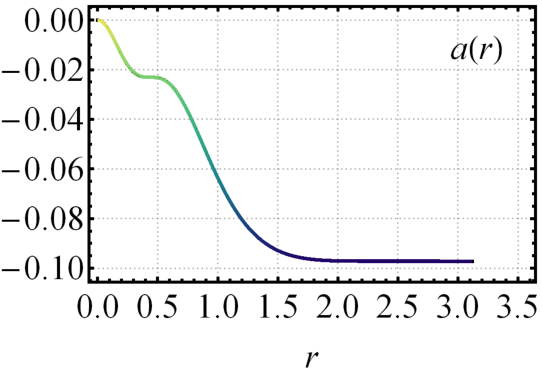} \hspace{0.5cm}%
	\includegraphics[width=5.cm]{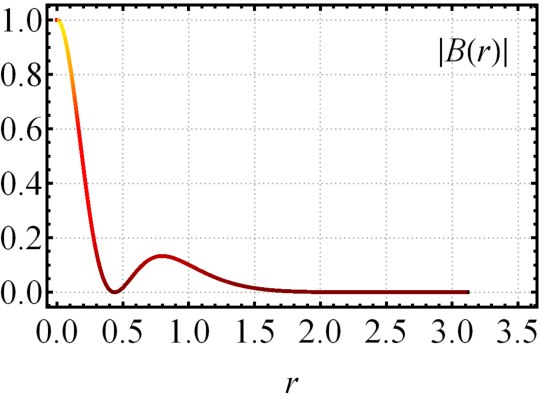} \hspace{0.5cm}%
	\includegraphics[width=5.cm]{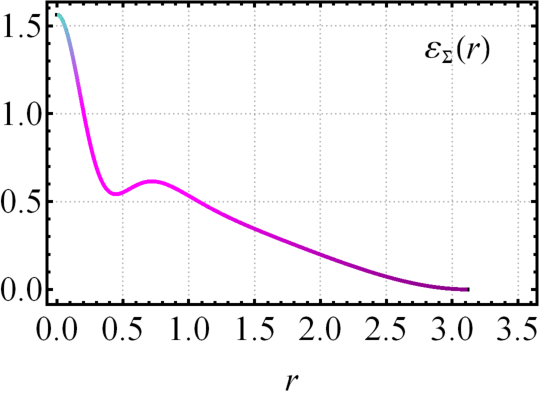}
	\includegraphics[width=5.3cm]{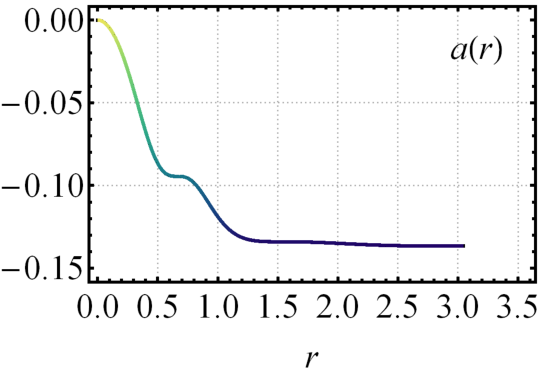} \hspace{0.5cm}%
	\includegraphics[width=5.cm]{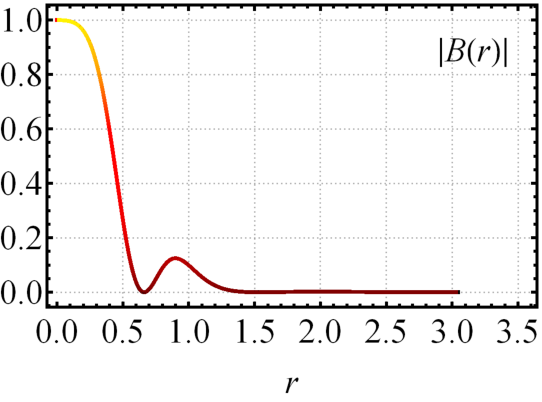} \hspace{0.5cm}%
	\includegraphics[width=5.cm]{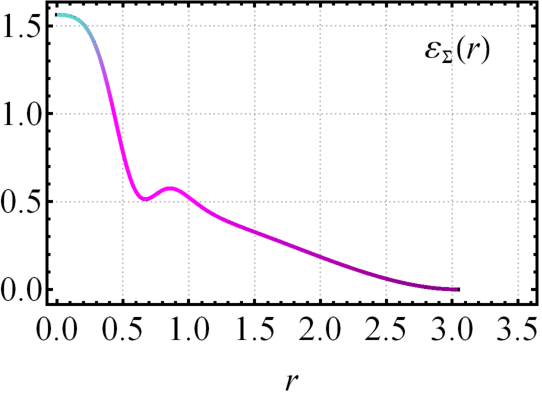}
	\caption{The gauge field profiles $a(r)$ (left panels),
		absolute value of the magnetic field $\vert B(r)\vert$ (middle panels) and	energy density $\protect\varepsilon_{\Sigma}(r)$ (right panels). We depict for $\protect\gamma=6 $ with $\protect\alpha=1$ (top) and $\protect\alpha=2$ (bottom).}
	\label{Fig2}
\end{figure}

\begin{figure}[tbp]
\includegraphics[width=5.2cm]{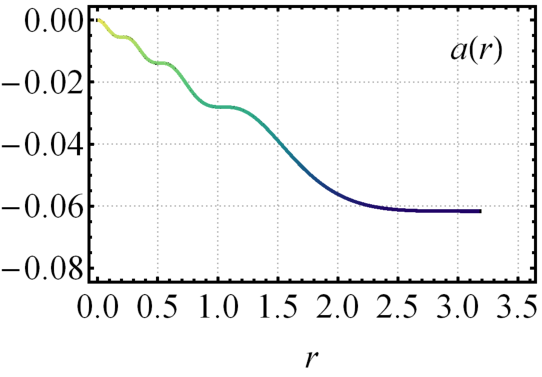}  \hspace{0.5cm}%
\includegraphics[width=5.cm]{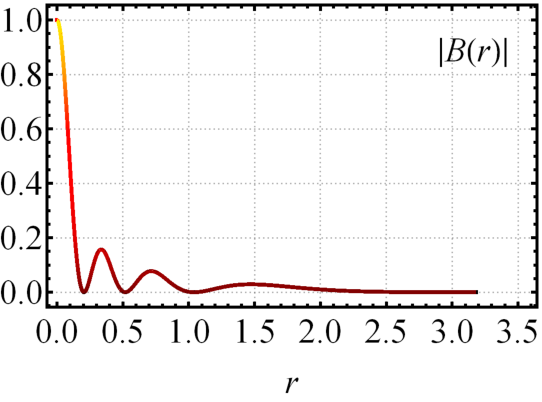}  \hspace{0.5cm}%
\includegraphics[width=5.cm]{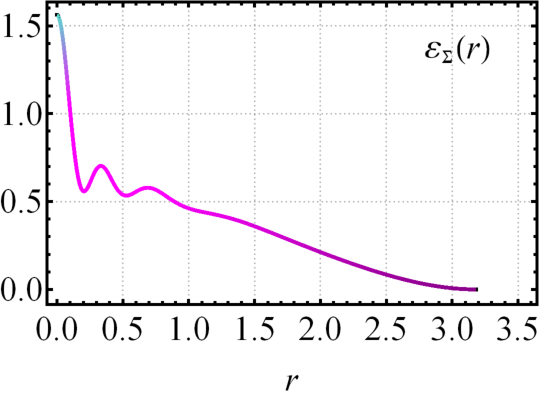}
\includegraphics[width=5.3cm]{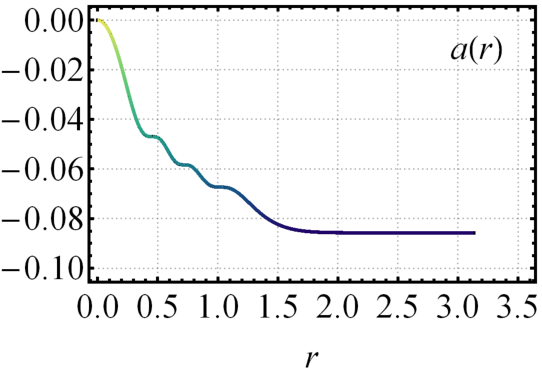}  \hspace{0.5cm}%
\includegraphics[width=5.cm]{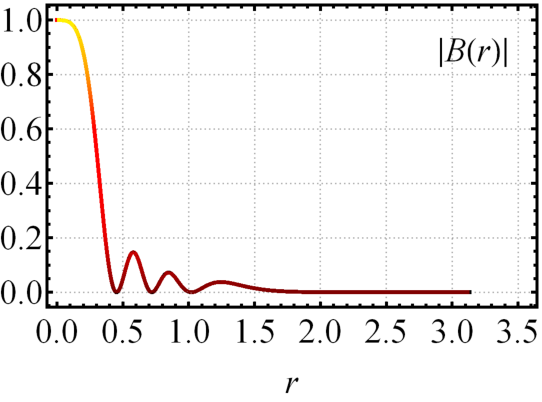}  \hspace{0.5cm}%
\includegraphics[width=5.cm]{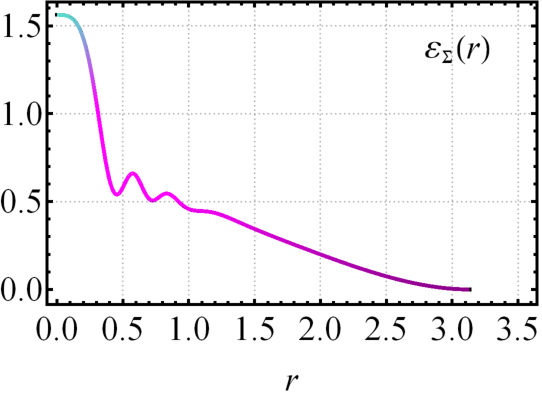}
\caption{The gauge field profiles $a(r)$ (left panels),
absolute value of the magnetic field $\vert B(r)\vert$ (middle panels) and
energy density $\protect\varepsilon_{\Sigma}(r)$ (right panels). We depict for $\protect\gamma=12 $ with $\protect\alpha=1$ (top) and $\protect\alpha=2$ (bottom).}
\label{Fig001}
\end{figure}

The field {profiles in the limit $r\rightarrow R$, i.e., when they reach the
corresponding vacuum value,} possess the following behavior:
\begin{equation}
h(r) \approx \frac{3^{2}}{2^{8}}\frac{W_{0}^{2}R^{2}}{\left( 1+a_{R}\right)
^{2}N^{2}}\left( r-R\right) ^{2}+\cdots
\end{equation}
\begin{equation}
a(r) \approx a_{R}+\frac{1}{2^{2}}\frac{g^{2}R^{2}}{\left( 1+a_{R}\right)
N^{2}}\mathcal{C}_{R}\left( r-R\right) ^{4}+\cdots \text{,}
\end{equation}%
where $\mathcal{C}_{R}$ is a constant given by

\begin{figure*}[t]
	\centering
	{\ \rule{0.\linewidth \includegraphics[width=6.35cm]{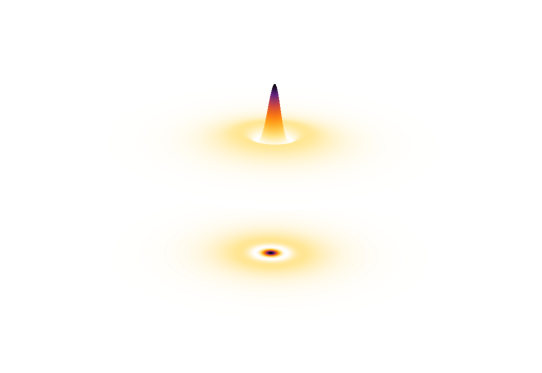}}{4cm}} {\
		\hspace{-0.82cm}\rule%
		{0.\linewidth \includegraphics[width=6.35cm]{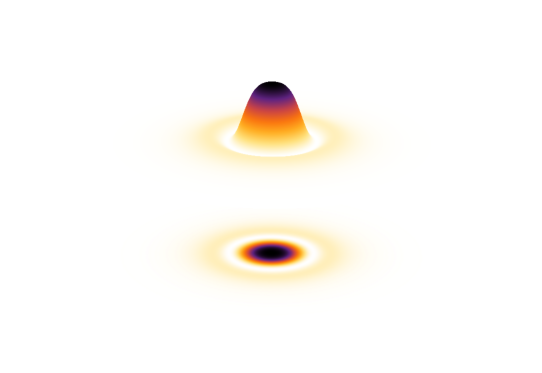}}{4cm}} {\ \hspace{%
			-0.8cm}\rule{0.\linewidth \includegraphics[width=6.35cm]{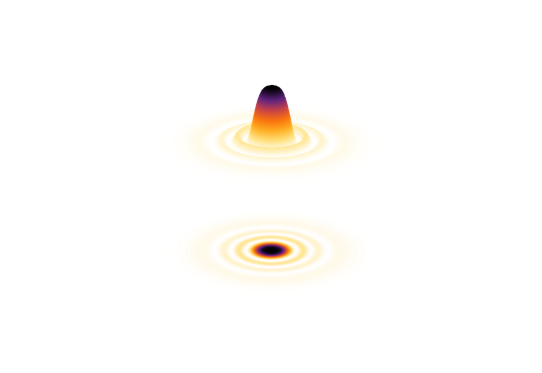}}{4cm}%
	}\vspace{-0.5cm}
	\caption{The magnetic field distributions (upper plots) with their
		corresponding projections in the plane (bottom plots). We depict for $%
		r_{0}=0.7$ with $\protect\gamma=6$ and $\protect\alpha=1$ (left plot), $%
		\protect\gamma=6$ and $\protect\alpha=2$ (middle plot), and $\protect\gamma %
		=12$ and $\protect\alpha=2$ (right plot).}
	\label{Fig002}
\end{figure*}

\begin{equation}
\mathcal{C}_{R}=\frac{3^{3}}{2^{12}}\frac{\lambda ^{2}W_{0}^{4}R^{2}}{\left(
1+a_{R}\right) ^{2}N^{2}}J_{0}^{2}\left( \frac{\gamma R^{\alpha }}{\sqrt{%
R^{2\alpha }+r_{0}^{2\alpha }}}\right) \text{,}
\end{equation}
depending on the parameters $\gamma$, $\alpha$ and $r_{0}$ that control the medium. Furthermore, the first relevant terms contributing to both the magnetic
field and energy density $\varepsilon _{\Sigma}$ in the limit $r\rightarrow R
$ are
\begin{equation}
B(r) =-\frac{g^{2}R}{\left( 1+a_{R}\right) N}\mathcal{C}_{R}\left(
r-R\right) ^{3}+\cdots
\end{equation}%
and%
\begin{eqnarray}
\varepsilon _{\Sigma } &=&\frac{3^{4}}{2^{12}}\frac{\lambda
^{2}W_{0}^{4}R^{2}}{\left( 1+a_{R}\right) ^{2}N^{2}}\left( r-R\right)
^{2}+\cdots\notag\\[0.2cm]
&& +\frac{3^{3}}{2^{12}}\frac{g^{2}\lambda ^{2}W_{0}^{4}R^{4}}{\left(
1+a_{R}\right) ^{4}N^{4}}\mathcal{C}_{R}\left( r-R\right) ^{6}+\cdots \text{.%
}
\end{eqnarray}

{We next} present the numerical solutions for the {\ fields by} solving the
system of BPS equations (\ref{bpsC11}) and (\ref{bpscII}){. Again we have
fixing} $N=1$, $W_{0}=1$, $\lambda =1$, $g=1$ and $r_{0}=1$ (when adopted
other value will be shown), {\ for different values of} $\alpha $ and $m$.

{To analyze how the compacton radius $R$ changes, we depict $R $ vs. $\alpha
$ (see left panel of Fig. \ref{Fig1}). We remark the existence of an
interval $0<\alpha \leq \bar{\alpha}$ (here $\bar{\alpha}\simeq {0.9}$)
where the compacton radius grows until to reach a maximum value $%
R_{\max}\simeq {3.12}$ when $\alpha =\bar{\alpha}$, a behavior unlike from
the previous case. Thereafter, for $\alpha > \bar{\alpha}$, the values of
the radius monotonically decrease tending asymptotically to a minimum value $%
R\rightarrow R_{\min}$ (here, $R_{\min}\simeq {2.51}$). This effect is
illustrated in the right panel of Fig. \ref{Fig1} where it is depicted $h(r)$
for some values of $\alpha>\bar{\alpha}$. The picture shows how the profiles
are modified, while the compacton radius decreases as $\alpha $ grows.}

{\ In Figs. \ref{Fig2} and \ref{Fig001} we have depicted the profiles for
the gauge field $a(r)$, magnetic field $B(r)$ and energy density $%
\varepsilon_{\Sigma}(r)$. They behave according to the analytical
expressions calculated previously at the boundaries. We also observe
plateaus arising along the gauge field profiles; the inner plateau that
shapes the core becomes bigger while $\alpha $ grows. Consequently, the
presence of the plateaus engenders ringlike structures in the profiles of
both the $B(r)$ and $\varepsilon _{\Sigma}(r)$. Thus, in Fig. \ref{Fig002},
we highlight the new effects in the magnetic field profiles: first, one
notices that $\alpha$ controls the core size, whose radius increases as $%
\alpha$ grows (see left and middle plots). Second, the parameter $\gamma$
controls the number of outer rings (surrounding the respective soliton's
core) that increase while $\gamma$ grows.}

\section{Conclusions}

\label{SecV}

We have shown that BPS skyrmions also arise in an enlarged restricted gauged
baby Skyrme model with $SO(3)\times \mathds{Z}_{2}$--symmetry. The $%
\mathds{Z}_{2}$--symmetry introduces a neutral field $\chi$ that includes
nonlinearities also engendering solitonic structures. The interaction
between both the gauge and scalar fields happens through the magnetic
permeability $\Sigma(\chi)$, i.e., via the coupling $\Sigma(\chi)F_{\mu\nu} F%
{^{\mu\nu}}$. The successful implementation of the BPS procedure allows for
obtaining the Bogomol'nyi bound and the corresponding self-dual equations
whose solutions saturate such a bound.

We have verified that the extended gauged baby Skyrme model proposed here
also engenders compact and noncompact topological structures. However, the
manuscript focuses on compact skyrmions by studying how they are affected by
including the $\mathds{Z}_{2}$--symmetry guiding the functional form of the
magnetic permeability $\Sigma(\chi)$. Then, to define the magnetic
permeability, we investigated two scenarios characterized by the
superpotentials engendering the $\chi^{4}$ and $\chi^{6}$ models,
respectively. Henceforth, the effects on the compactons immersed in those
media were analyzed separately. Both analyses reveal that the $\mathds{Z}_{2}
$--symmetry alters the profiles of the solitons. Among the features induced
by the $\mathds{Z}_{2}$--symmetry, we list: (i) it changes the compacton
radius size; (ii) it promotes the arising of the ringlike format, and (iii)
it controls the inner region size and the number of outer rings surrounding
the center.

We know the physical differences between the skyrmions studied here \cite%
{fnote1} and the magnetic skyrmions researched in condensed matter,
including those obtained considering the DM interaction. Both cases possess
finite energy topological configurations, which include compactons. Then, a
natural question arises about the possibility of a unified description by
investigating whether both interactions could coexist to engender novel
solutions. Indeed, in a recent study reported in Ref. \cite{Funa1}, the
authors introduce the DM interaction in the baby Skyrme model through the
effective potential technique, even obtaining compact skyrmions. This
approach may pave the way to understanding the connection between the
skyrmion studied here and the magnetic skyrmions arising in condensed
matter. Still in this direction, it could be possible to initiate the study
concerning the BPS structure of the baby Skyrme model in the presence of the
DM interaction.

Finally, already we have some issues under consideration, including the baby
Skyrme model in the presence of the DM interaction, as the engendering of
compact skyrmions (carrying only magnetic flux and the ones with both
magnetic flux and electrically charged) in the presence of magnetic media,
magnetic impurities, and other topological (nontopological) objects.
Advances in these directions we will report elsewhere.

\begin{acknowledgments}
This study was financed in part by the Coordena\c{c}\~ao de Aperfei\c{c}oamento de Pessoal de N\'{\i}vel Superior - Brasil (CAPES) - Finance Code
001. We thank also the Conselho Nacional de Desenvolvimento Cient{\'\i}fico
e Tecnol\'ogico (CNPq), and the Funda\c{c}\~ao de Amparo \`a Pesquisa e ao
Desenvolvimento Cient{\'\i}fico e Tecnol\'ogico do Maranh\~ao (FAPEMA)
(Brazilian Government agencies). R. C. acknowledges the support from the
grants CNPq/306724/2019-7, FAPEMA/Universal-01131/17,
FAPEMA/Universal-00812/19, and FAPEMA/APP-12299/22.  In particular, A. C. S. thank the grants CAPES/88882.315461/2019-01 and CNPq/150402/2023-6 and C. A. I. F. thanks the support from FAPEMA/BD-05890/23.
\end{acknowledgments}


\begin{thebibliography}{99}
\bibitem{fnote} We use this notation for the Skyrme field and the unitary
vector along the manuscript; besides, the quantity $\hat{n} \cdot \vec{\phi}$
will be denoted as $\phi_{n}$.
\bibitem{fnote1} Here, it is worthwhile to point out that these solutions of
the restricted gauged baby Skyrme model are not the magnetic skyrmions
describing the material magnetization in condensed matter systems.
\bibitem{metrica} Here, for the $(1+2)$-dimensional Minkowski space we adopt
the metric with signature $\left( +,-,-\right)$. Moreover, we take the
natural units $\hbar =c=1$.
\end{thebibliography}
\end{document}